\documentclass[aps,prc,amssymb,amsmath,superscriptaddress,reprint]{revtex4-1}

\usepackage[english]{babel}
\usepackage[utf8]{inputenc}

\usepackage[dvips]{graphicx}
\usepackage[breaklinks]{hyperref}
\usepackage[sort&compress]{natbib}

\begin{document}

\title{Baryon Resonances in a Chiral Hadronic Model for the QCD
  Equation of State}

\author{Philip Rau}%
\email[]{rau@th.physik.uni-frankfurt.de}%
\affiliation{Institut f\"ur Theoretische Physik, Johann Wolfgang
  Goethe-Universit\"at, Max-von-Laue-Str.\ 1, 60438 Frankfurt am Main,
  Germany}%
\affiliation{Frankfurt Institute for Advanced Studies (FIAS),
  Ruth-Moufang-Str.\ 1, 60438 Frankfurt am Main, Germany}%

\author{Jan Steinheimer}%
\affiliation{Institut f\"ur Theoretische Physik, Johann Wolfgang
  Goethe-Universit\"at, Max-von-Laue-Str.\ 1, 60438 Frankfurt am Main,
  Germany}%
\affiliation{Frankfurt Institute for Advanced Studies (FIAS),
  Ruth-Moufang-Str.\ 1, 60438 Frankfurt am Main, Germany}%

\author{Stefan Schramm}%
\affiliation{Institut f\"ur Theoretische Physik, Johann Wolfgang
  Goethe-Universit\"at, Max-von-Laue-Str.\ 1, 60438 Frankfurt am Main,
  Germany}%
\affiliation{Frankfurt Institute for Advanced Studies (FIAS),
  Ruth-Moufang-Str.\ 1, 60438 Frankfurt am Main, Germany}%

\author{Horst St\"ocker}%
\affiliation{Institut f\"ur Theoretische Physik, Johann Wolfgang
  Goethe-Universit\"at, Max-von-Laue-Str.\ 1, 60438 Frankfurt am Main,
  Germany}%
\affiliation{GSI Helmholtzzentrum f\"ur Schwerionenforschung GmbH,
  Planckstr.\ 1, 64291 Darmstadt, Germany}%


\begin{abstract}
  In this paper we study the influence of hadronic resonances on the
  phase diagram calculated with an effective chiral flavour SU(3)
  model. We show that varying the couplings of the baryonic resonances
  to the attractive scalar and the repulsive vector fields has a major
  impact on the order and location of the chiral phase transition and
  the possible existence of a critical end point as well as on the
  thermodynamic properties of the model. Furthermore, we study
  (strange) quark number fluctuations and show the related
  susceptibilities both at zero baryochemical potential and when
  crossing the phase transition line at three different points in the
  $T$--$\mu$ plane. We obtain the best agreement with current lattice
  data if we choose a rather strong vector coupling which in our model
  limits the phase transition to a smooth crossover and implies the
  non-existence of a critical end point.
\end{abstract}

\maketitle
\section{Introduction}
\label{sec:introduction}

Currently there are many efforts both on the theoretical, as well as
on the experimental side to gain knowledge about the phase diagram of
strongly interacting nuclear matter. Experiments with highly energetic
colliding gold nuclei at the Relativistic Heavy Ion Collider (RHIC)
suggest that a new state of matter, comparable to a nearly perfect
fluid, is created at high temperatures~\cite{Arsene2005, Back2005,
  Adams2005, Adcox2005}. While these experiments at RHIC and the Large
Hadron Collider (LHC), performed at very high beam energies, aim at
the high-temperature, low-density region of the phase diagram, there
are also experiments in which higher baryonic densities may be
reached, as for example planned at the upcoming Facility for
Antiproton an Ion Research (FAIR) at GSI. Common to all these
experiments is the search for the phase transition of strongly
interacting matter from a confined state of hadrons at low
temperatures to a state with deconfined quarks and gluons (QGP) as it
is predicted for high densities and high
temperatures~\cite{Gyulassy2005}. This phase transition is commonly
referred to as deconfinement transition~\cite{Svetitsky1982}. The
symmetries of quantum chromodynamics (QCD), however, imply another,
\emph{chiral} phase transition at which the chiral symmetry is
restored and the masses of the baryons, or constituent quarks,
vanish~\cite{Nambu1961, Kirzhnits1972, Weinberg1974}. For this
transition, on which we will mainly focus our studies, the chiral
condensate $\sigma$ acts as the corresponding order parameter.\par
Since QCD can not be treated perturbatively at low temperatures, most
information we have about the phase diagram of QCD matter today comes
from lattice QCD calculations~\cite{Karsch2000, Fodor2002, Fodor2003,
  Fodor:2004nz, Karsch2005, Aoki2006, Cheng2006, Cheng2007, DeTar2008,
  Petreczky2009}. In the region of vanishing baryochemical potentials
$\mu = 0$ and finite temperature, lattice QCD yields reliable results
that show a smooth crossover transition~\cite{Aoki2006}. For finite
potentials $\mu > 0$, however, results from lattice QCD are not
available because of the failing of standard Monte Carlo sampling
methods due to the so-called sign-problem. Currently there are various
methods to extend the lattice results from $\mu = 0$ to the region of
small chemical potentials~\cite{Fodor:2001pe, Fodor2002, Fodor2007,
  Allton:2002zi, Forcrand2003, D'Elia2003, D'Elia2007, Allton2005,
  Fodor2003, Forcrand2008, Kaczmarek2011,Wu2007}. Some lattice QCD
groups suggest that the phase transition becomes first order at a
critical endpoint~\cite{Stephanov1998, Stephanov1999, Fodor:2001pe,
  Fodor:2004nz, Stephanov2006} with its coordinates varying
considerably for different calculations. Other more recent
results~\cite{Endrodi:2011gv} favour the exclusive existence of a
crossover transition for a wide range of baryon densities, implying
the non-existence of such a critical point. Moreover, the exact
position of the phase transition in the phase diagram is subject of a
lively and ongoing debate.\par
Other common theoretical approaches to QCD are effective models to
study specific properties of QCD matter~\cite{Walecka1974, Serot1986,
  Boguta1977, Boguta:1981px, Khvorostukhin2007, Khvorostukhin2008,
  Nambu1961, Nambu1961b}. In our approach we will use an effective
model for the QCD equation of state to study the properties of the
phase diagram of nuclear matter. Our model is able to reproduce the
well known saturation properties of nuclear
matter~\cite{Papazoglou:1997uw, Papazoglou:1998vr}.\par
The degrees of freedom in our model are all known baryons and baryonic
resonances up to masses of $m = 2.6$~GeV. This approach resembles an
interacting hadron resonance gas (HRG). The HRG is known to give a
good description of thermodynamic quantities at low temperatures ($T
\le T_c$) and was often used to reproduce lattice QCD results for
thermodynamics and fluctuations of conserved charges at low
temperatures~\cite{Karsch2003, Karsch2003a, Karsch2005, Huovinen2010,
  Borsanyi2010, Borsanyi2010a, Cheng2009}. The particle production in
heavy ion collisions may also be reasonably well described with use of
the HRG~\cite{Cleymans1998, Cleymans1999, Becattini2001,
  Braun-Munzinger2001, Andronic2003}. Therefore, our first approach to
use only hadronic degrees of freedom seems to be appropriate up to at
least temperatures in the range of $T_c$. For higher temperatures, at
the latest when the deconfinement phase transition sets in, this
approach breaks down and, in a future step, partonic degrees of
freedom need to be taken into account as it was done in
Refs.~\cite{Steinheimer:2009hd, Steinheimer:2010ib} for example.\par
This paper is organised as follows. First we introduce the chiral
$SU(3)$ model in section II. In section III, we present results for
the order parameter of the chiral condensate $\sigma$ at zero and
nonzero baryochemical potentials, thermodynamic quantities from the
model, and quark susceptibilities at different points in the phase
diagram. Susceptibilities are interesting quantities since they
resemble fluctuations of conserved charges which itself are closely
linked to phase transitions. They offer an effective possibility of
comparison between results from theory and experiment because the
susceptibilities can be related to measured fluctuations of particle
production. Of particular interest are susceptibility ratios since
they are not dependent on the volume or the impact parameter of the
underlying system~\cite{Koch2008}. This work closes with a conclusion
in section IV.

\section{Model}
\label{sec:model}

In our model, a $SU(3)$-flavour sigma-omega model using the non-linear
realization of chiral symmetry (see Refs.~\cite{Boguta:1981px,
  Papazoglou:1997uw, Papazoglou:1998vr, Dexheimer:2008ax} for a
comprehensive review), the Lagrangian in mean field approximation has
the form
\begin{equation}
  \label{eq:lagrangian_summary}
  \mathcal{L} = \mathcal{L}_{\rm kin} + \mathcal{L}_{\rm int} +
  \mathcal{L}_{\rm meson}.
\end{equation} 
Here, the first term represents the kinetic energy of the hadrons, the
terms
\begin{equation}
  \label{eq:L_int}
  \mathcal{L}_{int} = -\sum_i \bar{\psi_i} \left( m^*_i + g_{i\omega}
    \gamma_0 \omega^0 + g_{i\phi}
    \gamma_0 \phi^0 \right) \psi_i\ 
\end{equation}
describe the interaction of the baryons with the scalar mesons
$\sigma$, $\zeta$ (attractive interaction, see
Eq.~(\ref{eq:effective_mass})) and the vector mesons $\omega$, $\phi$
(repulsive interaction) respectively. The summation index $i$ runs
over the baryon octet ($N$, $\Lambda$, $\Sigma$, $\Xi$), the baryon
decuplet ($\Delta$, $\Sigma^*$, $\Xi^*$, $\Omega$), and all heavier
resonance states up to masses of $m_{N^*} = 2600$~MeV. We only include
hadronic resonances whose existence is considered to be very likely
according to the Particle Data Group listings~\cite{Nakamura2010}
where they are recorded with a minimum three-star rating. Since the
listed masses of some heavy resonances may cover a broad range, all
particles are included with their average mass.\par
The third term of the Lagrangian
\begin{align}
  \mathcal{L}_{\rm meson} &= \mathcal{L}_{\rm vec} +\mathcal{L}_{0} +
  \mathcal{L}_{\rm ESB}\\
  &= + \frac{1}{2} \left( m^2_{\omega}
    \omega^2 + m^2_{\phi} \phi^2 \right) \notag \\
  & \quad +g_4 \left( \omega^4 + \frac{\phi^4}{4} + 3\omega^2 \phi^2 +
    \frac{4 \omega^3 \phi}{\sqrt{2}} +
    \frac{2 \omega \phi^3}{\sqrt{2}} \right) \notag \\
  &\quad - \frac{1}{2} k_0 (
  \sigma^2 + \zeta^2) + k_1(\sigma^2 + \zeta^2)^2   \label{eq:L_meson} \\
  &\quad + k_2\left(\frac{\sigma^4}{2} + \zeta^4\right) +
  k_3\sigma^2\zeta + k_4
  \ln{\frac{\sigma^2\zeta}{\sigma_0^2\zeta_0}} \notag \\
  &\quad - m_\pi^2 f_\pi\sigma+\left(\sqrt{2}m_k^ 2f_k \notag
    -\frac{1}{\sqrt{2}}m_\pi^ 2 f_\pi\right)\zeta
\end{align}
includes the self interactions of the vector mesons and the scalar
mesons together with the last two terms describing the explicit
symmetry breaking.\par
The effective masses of the baryons
\begin{equation}
  \label{eq:effective_mass}
  m_{i}^* = g_{i\sigma}\sigma + g_{i\zeta}\zeta + \delta m_i
\end{equation}
are created by the coupling of the baryons to the scalar meson fields
(i.e.\ the non-strange chiral condensate $\sigma$ and its strange
equivalent $\zeta$), together with an explicit mass of at least
$\delta m_N = 150$~MeV. In this way at high temperatures and baryonic
densities the decreasing $\sigma$-field leads to smaller baryon masses
and thus to the restoration of chiral symmetry. Thereby the nucleons
which have the smallest explicit mass have lost roughly $45\%$ of
their vacuum mass at $T_c$. Since our model only includes hadronic
degrees of freedom it can only be applied in the hadronic regime up to
temperatures slightly above $T_c$. This also prevents baryonic masses
from getting to small so that the mean field approximation should
remain valid and no fluctuations need necessarily to be taken into
account as it was for example done in Refs.~\cite{Mocsy2004,
  Bowman2009}.\par
The effective masses of the pseudoscalar mesons and vector mesons are
given by the second derivative of the mesonic potential
$\mathcal{V}_{\rm meson} = - \mathcal{L}_{\rm vec} - \mathcal{L}_{0} -
\mathcal{L}_{\rm ESB}$ with respect to the respective mesons $\xi_j$
at the minimum of the grand canonical potential (see
Eq.~\eqref{eq:grand_canon_pot})~\cite{Zschiesche2004}
\begin{align}
  \label{eq:effective_meson_mass}
  m^*_j &= \frac{\partial^2}{\partial \xi^2_j} \mathcal{V}_{\rm
    meson}(\zeta^0_j),\\
  \xi_j &= \pi, \eta, \eta', K, \bar{K};\; \rho, \omega, \varphi, K^*,
  \bar{K}^*. \nonumber
\end{align}\par
The couplings of the baryon octet to the mesonic fields and the
mesonic potential are chosen in such a way as to reproduce the
well-known vacuum masses, the nuclear ground state properties (e.g.\
the correct binding energy), and the asymmetry energy. The coupling
strengths of the baryons of the decuplet and all heavier resonances
are scaled to the nucleon couplings via the parameters $r_s$, $r_v$
according to
\begin{align}
  \label{eq:couplings}
  g_{B \sigma, \omega} &= r_{s,v} \cdot g_{N \sigma, \omega},\\
  g_{B \zeta, \phi}  &= r_{s,v} \cdot g_{N \zeta, \phi}. 
\end{align}
In this paper we systematically analyse the influence of the baryonic
resonances on the hadronic matter properties. Therefor we adjust the
vector coupling parameter $r_v$, controlling the abundance of the
baryonic resonances at finite baryochemical potentials in the model,
in order to study the influence on the resulting phase diagram and the
thermodynamic properties of the model. We use this one-parameter
approximation in order not to be swamped by a plethora of of unknown
coupling constants of the various hadronic multiplets. In this work we
set the scalar coupling parameter fixed to $r_s = 0.97$ which ensures
a smooth crossover phase transition at zero baryochemical potential
(see Fig.~\ref{fig:sigma}~(b) for a study of the effect of $r_s$). In
general the scalar couplings are fixed by reproducing the particles'
vacuum masses (except for the explicit mass term $\delta m_i$).\par
As a reference we also perform calculations for an ideal HRG that is
not interacting with the mesonic fields. In this particular case all
baryon couplings to the fields $g_{B}$ are set to zero and the masses
of all particles are fixed at their tabulated vacuum expectation
value.\par
The grand canonical potential of our model takes the form
\begin{equation}
  \label{eq:grand_canon_pot}
  \frac{\Omega}{V} = -\mathcal{L}_{\rm int} - \mathcal{L}_{\rm meson}
  + \Omega_{\rm th}
\end{equation}
with the thermal contribution of the hadrons in the model
\begin{align}
  \label{eq:gc_pot_thermal}
  \Omega_{\rm th} = &- T \sum\limits_{i\in B} \frac{\gamma_i}{(2
    \pi)^3} \int d^3k \; \left( \ln \left[ 1 + e^{ -\frac{1}{T} \left(
          E^*_i(k) - \mu^*_i \right)} \right]\right. \nonumber \\
    &\hspace{5em}\left. +\ln \left[ 1 + e^{ -\frac{1}{T} \left(
          E^*_i(k) + \mu^*_i \right)} \right] \right)\\
  &+ T \sum\limits_{j\in M} \frac{\gamma_j}{(2
    \pi)^3} \int d^3k \; \ln \left[ 1 - e^{ -\frac{1}{T} \left(
          E^*_j(k) - \mu_j\right)} \right]. \nonumber
\end{align}
Here the sums run over all baryons $B$ (anti baryons are explicitly
included in the second term of the first integral) and all mesons $M$
in the model, $\gamma_{i,j}$ stands for the spin-isospin-degeneracy
factor of the respective particle species $i, j$ and $E^*_{i,j}(k) =
\sqrt{k^2 + m_{i,j}^{*2}}$ for the single particle energies. The
effective baryochemical potential is defined as $\mu^*_i = \mu_i -
g_{i \omega} \omega - g_{i \phi} \phi$. The grand canonical potential
then leads to the thermodynamic quantities of the system, i.e.\ the
pressure $p$ and the energy and entropy density $e$, $s$, together
with the densities of the particular particle species $\rho_i$.\par
Another effect we include in our model (following the works of
Refs.~\cite{Rischke1991, Steinheimer:2010ib}) is the effective
suppression of baryonic states at high densities due to excluded
volume effects~\cite{Baacke1977, Hagedorn1980, Gorenstein1981,
  Hagedorn1983}.  While so far all particles were regarded as point
like, we now can introduce for every single particle $j$ an average
finite volume $v_{\rm ex}^j$ excluded from the total volume of the
system $V$. In a first and very basic consideration this excluded
volume in a non-relativistic definition takes the form
\begin{equation}
  \label{eq:excluded_volume}
  v_{\rm ex}^j = \frac{1}{2\,a_i}\; \frac{4}{3} \pi (2\,r)^3,
\end{equation}
where $r$ is the mean radius of all particles in the model and the
parameter $a_i$ defines the excluded volume for the different particle
species $i$. In our case we set $a_{\rm B, M} = 1$ as a first
approximation for all baryons and mesons. This modification leads to
the altered chemical potential
\begin{equation}
  \label{eq:chem_pot_new}
  \mu_j' = \mu_j - v_{\rm ex}^j P,
\end{equation}
where $P$ is the sum over all partial pressures. All thermodynamic
observables must now be expressed in terms of $T$ and
$\mu_j'$. Furthermore, a volume correction factor for each particle
species
\begin{equation}
  \label{eq:vol_correction_factor}
  f_i = \frac{V_i'}{V} = \left( 1 + \sum \limits_j v_{\rm ex}^j \rho_j
  \right)^{-1}
\end{equation}
with $V_i'$ the volume not being occupied, has to be introduced in
order to express the densities in a thermodynamically consistent way
\begin{align}
  \label{eq:new_densities}
  \rho_i' &= f_i \rho_i,\\
  e' &= \sum \limits_i f_i e_i,\\
  s' &= \sum \limits_i f_i s_i.
\end{align}
The inclusion of the excluded volume effects also has an impact on the
nuclear ground state properties and thus this approach is considered
only as a first attempt to study the impact of the suppression of
baryons on the phase transition and on the thermodynamic properties of
the model. However, reasonable nuclear ground state properties can be
achieved by recalibration of the parameters as it was shown in
Ref.~\cite{Steinheimer2011a}.

\section{Results}
\label{sec:results}

\subsection{Order Parameters}
\begin{figure}[t!]
  \centering 
  \includegraphics[width=1.\columnwidth]{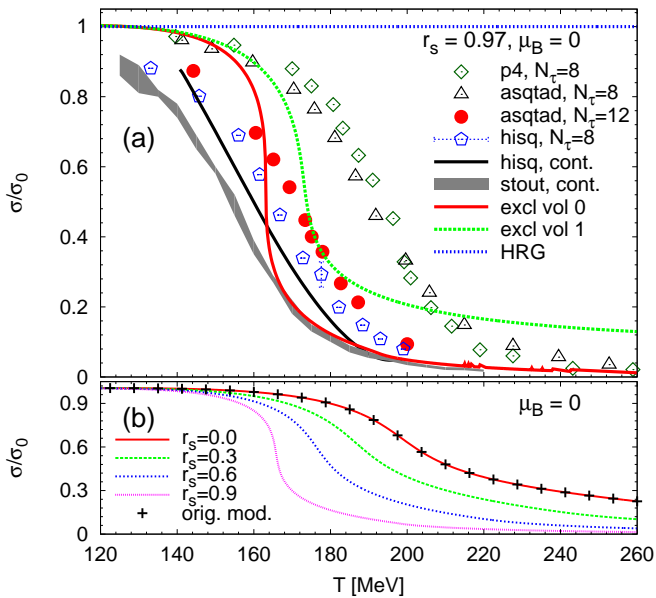}
  \caption{(Color online) (a) Normalised order parameter for the
    chiral condensate $\sigma/\sigma_0$ as a function of $T$ at
    $\mu_B=0$ with (green dotted line) and without (red solid line)
    excluded volume corrections compared to lattice data. The blue
    dotted line at $\sigma/\sigma_0 = 1$ shows the reference line of
    the non-interacting ideal hadron resonance gas. Here and in the
    following plots the lattice data are shown as calculated with the
    asqtad~\cite{Bazavov2009a}, the
    hisq~\cite{Bazavov2010e,Bazavov2010}, the
    p4~\cite{Cheng2009,Bazavov2009a,Cheng2010}, and the stout
    action~\cite{Borsanyi2010,Borsanyi2010b,Aoki2009} on lattices with
    different temporal extent $N_{\tau}$. Panel (b) shows
    $\sigma/\sigma_0$ for different strength of the scalar coupling
    $r_s$ together with results from our model without any resonance
    states.}
  \label{fig:sigma}
\end{figure}

We start our investigation of the model properties with the
calculation of the order parameter for the chiral transition $\sigma$
at zero baryochemical potential $\mu_B$. Figure~\ref{fig:sigma}~(a)
shows the normalised chiral order parameter as a function of the
temperature together with data from lattice QCD calculations. Here and
in the following, the lattice data we refer to were obtained by
different collaborations using various lattice actions (asqtad, hisq,
p4 and stout) and temporal spacings of the lattice ($N_{\tau} = 4 -
12$)~\cite{Bazavov2009a, Bazavov2010e, Bazavov2010, Cheng2009,
  Bazavov2009a, Cheng2010, Borsanyi2010, Aoki2009, Ejiri2004,
  Endrodi:2011gv}.
\par
With the scalar couplings fixed to $r_s = 0.97$ we obtain a smooth
crossover in $\sigma$ for both the model with excluded volume (green
dashed line) effects due to finite size of the particles and without
it (red solid line). The critical temperature, defined as the point
with the largest increase in $\sigma$, is found to be $T_c = 164$~MeV
without the excluded volume effect and $T_c = 174$~MeV if the finite
size effects are taken into account. The results for $\sigma$ without
the excluded volume effects are in qualitatively good agreement with
lattice QCD calculations which predict a smooth crossover at $\mu_B =
0$ and critical temperatures in the range from $T_c = 155$~MeV to
$200$~MeV~\cite{Aoki2006, Bazavov2009a, Bazavov2010, Bazavov2010e,
  Aoki2009, Cheng2010, Borsanyi2010, Detar2007, Endrodi:2011gv}, where
the newest continuum extrapolated data from the Hot-QCD and the
Wuppertal-Budapest Collaborations predict critical temperatures close
to $T = 160$~MeV consistently. For temperatures below $T_C$ the slope
of the chiral condensate from our model is relatively steep and
deviates from those suggested by lattice QCD calculations. This is
mainly due to neglecting contributions from pseudoscalar meson self
interactions which are important at low
temperatures~\cite{Gerber1989}. The slope of $\sigma$ could be leveled
by including the $\pi$ self interaction in the
model~\cite{Steinheimer2011a, Mishra2008}.\par
The impact of the heavy resonances states can be seen from
Fig.~\ref{fig:sigma}~(b) where the black dots show the normalised
chiral order parameter as a function of the temperature as calculated
with our original model which does not include any hadronic resonance
states~\cite{Papazoglou:1997uw, Papazoglou:1998vr}. It shows that
without the hadronic resonance states the transition is much smoother
and exhibits a higher critical temperature close to $T_c =
200$~MeV. The same effect is achieved by decreasing the strength of
the attractive coupling of the baryons to the scalar
$\sigma$-field. This behavior is also shown in
Fig.~\ref{fig:sigma}~(b) for values of $r_s$ from $0$ to $0.9$. The
weaker the scalar coupling is, the flatter is the transition and the
critical temperature moves to higher temperatures. For vanishing
scalar couplings (red line) the results are the same as in the
original model. Note that for small values of $r_s$ one would have to
introduce large values for the explicit mass term as the sum of the
meson-field generated mass and the explicit term have to reproduce the
vacuum masses of the states.\par
Next we extend this study of $\sigma$ to nonzero baryochemical
potentials, i.e.\ the whole $T$--$\mu_B$ plane. One of the major
interests in the study of the phase diagram of strongly interacting
matter is certainly to obtain information about the chiral and
deconfinement phase transitions and the search for a possible critical
end point which divides the crossover phase transition at vanishing
baryochemical potential from a first order phase transition at finite
chemical potentials. Since lattice QCD calculations are systematically
limited to $\mu_B = 0$ due to the so-called sign problem, different
lattice QCD groups use various methods to extend their results to
non-vanishing potentials. Unfortunately, this did not lead to a
consistent picture for the critical end point until now. While results
from (2+1)-flavour QCD calculations suggested the critical end point
at a critical quark chemical potential in the range from $\mu_B^{\rm
  crit} \approx 725$~MeV~\cite{Fodor:2001pe} to $\mu_q^{\rm crit}
\approx 140$~MeV~\cite{Ejiri2004} (with $3\mu_q = \mu_B$) and
$\mu_B^{\rm crit} \approx 360$~MeV~\cite{Fodor:2004nz}, more recent
studies of the Wuppertal-Budapest Collaboration~\cite{Endrodi:2011gv}
doubt the existence of a critical end point at all and suggest a broad
crossover phase transition over the whole $T$--$\mu_B$ plane.\par
\begin{figure}[t]
  \centering 
  \includegraphics[width=1.\columnwidth]{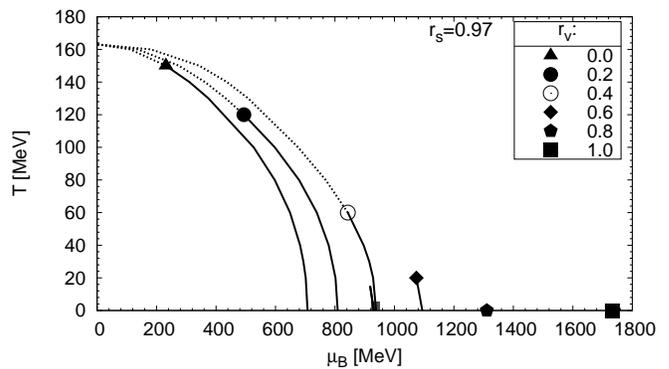}
  \caption{Phase transition lines of the chiral order parameter
    $\sigma$ in the $T$--$\mu_B$ plane for different values of $r_v$
    together with the nuclear ground state at $\mu_B = 939$~MeV, $T =
    0$ and the liquid gas first order phase transition. The point on
    each phase transition line depicts the critical end point,
    separating a first order phase transition (solid lines) from a
    crossover (doted lines). Note that for values $r_v > 0.6$ there is
    solely a very broad crossover in the whole $T$--$\mu_B$ plane.}
  \label{fig:phase_diagram}
\end{figure} 
In Fig.~\ref{fig:phase_diagram} we show the results from our model
without excluded volume effects for different values of the coupling
constant of the repulsive vector interaction $r_v$. In this figure the
lines represent the biggest increase of the $\sigma$-field when varying
$T$ and $\mu$. While a discontinuous first order phase transition in
$\sigma$ is depicted by a solid line, a smooth crossover in $\sigma$
is drawn with a dotted line. The point on each phase transition line
stands for the critical end point for the specific choice of $r_v$. We
find that increasing the vector coupling strength leads to a stronger
suppression of heavier particles and thus to a phase transition at
higher chemical potentials. Note that for all coupling strengths
smaller than $r_v = 0.4$ the nuclear ground state at $\mu_B =
939$~MeV, $T = 0$ and the first order liquid gas phase transition are
located above the phase boundary in the chirally restored
phase. Therefore, within our simple model approximations those small
vector couplings do not lead to physically reasonable solutions. Thus,
we conclude according to our model the smallest critical chemical
potential (for $r_v = 0.4$) is located at $\mu_B^{\rm crit} \approx
840$~MeV which is two times higher than the value suggested in
Ref.~\cite{Ejiri2004}.\par
For vector couplings of $r_v = 0.6$, in our model, there is a very
small temperature range of first order phase transition going up to
$T^{\rm crit} \approx 20$~MeV, at higher temperatures there is a
broad-ranged crossover transition. For all higher vector couplings,
the first order phase transitions vanish completely and only a broad
crossover region remains.

\subsection{Thermodynamics}
\begin{figure}[tb]
  \centering
  \includegraphics[width=1.\columnwidth]{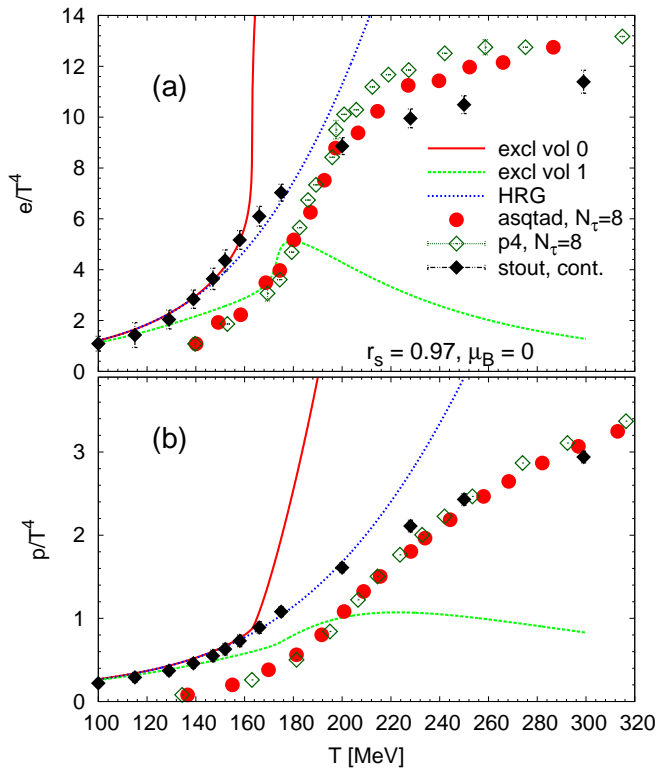}
  \caption{(Color online) Energy density (a) and the pressure (b)
    divided by $T^4$ at $\mu_B= 0$ as a function of the
    temperature. Depicted are results for the interacting hadronic gas
    with (green dashed line) and without (red solid line) excluded
    volume corrections and for the non-interacting ideal hadron
    resonance gas (blue dotted line) together with lattice data.}
  \label{fig:thermodyn_quantities}
\end{figure}
The thermodynamic quantities, as predicted by our model at $\mu_B = 0$
are depicted in Fig.~\ref{fig:thermodyn_quantities}. Panel (a) shows
the energy density and panel (b) shows the pressure both divided by
$T^4$, in comparison to lattice data. Here and in the following plots,
the solid red curve represents the results from our regular model
including all baryonic resonances, the green dashed curve depicts the
results from the model with the excluded finite size volume effects
taken into account and the blue dashed line stands for the
non-interacting ideal HRG without finite size effects.\par
For the standard case in our model, at $T_c$ the energy density and
the pressure rise rapidly due to the emerging abundance of particles
in the system because of their decreasing masses. Besides a slightly
higher critical temperature, the suppression of heavier particles
caused by the finite size effects leads to a much smaller maximum of
the energy density at $T_C$ and a decrease for higher temperatures. In
this case, three times the pressure divided by $T^4$ exhibits only a
slight increase with a maximum around $T_c$ and decreases smoothly
again for higher temperatures. The HRG results rise monotonically, as
expected.\par
\begin{figure}[tb]
  \centering
  \includegraphics[width=1.\columnwidth]{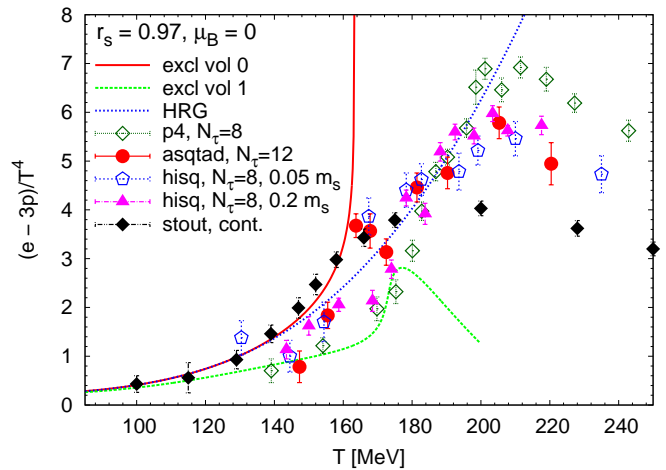}
  \caption{(Color online) The interaction measure, three times the
    pressure subtracted from the energy density over $T^4$, as a
    function of the temperature at $\mu_B = 0$ for the interacting
    hadronic gas with (green dashed line) and without (red solid line)
    excluded volume corrections and for the non-interacting ideal
    hadron resonance gas (blue dotted line) compared to lattice
    data~\cite{Aoki2009,Bazavov2009a,Petreczky2009,Cheng2009,
      Cheng2010,Bazavov2010e,Bazavov2010a,Borsanyi2010,Borsanyi2010b}. The
    data for the hisq action is shown for two different values of the
    light quark mass $m_l = 0.05\, m_s$ and $0.20\, m_s$, where $m_s$
    stands for the strange quarks mass.}
  \label{fig:int_measure}
\end{figure}
Figure~\ref{fig:int_measure} shows the so-called interaction measure
defined as the energy density minus three times the pressure divided
by $T^4$ as a function of the temperature. Again, the results from our
regular model (red curve) show a rapid increase of this quantity at
$T_c$, the HRG rises monotonously being in qualitatively good
agreement in the region around the critical temperature with lattice
data~\cite{Bazavov2009a, Cheng2009, Cheng2010} which shows a peak
slightly above $T_c$.\par
Comparing our results for the thermodynamic quantities to lattice
data, we state that a good agreement up to the critical temperature is
achieved for our results from the interacting HRG without excluded
volume effects and the continuum extrapolated stout action with
physical quark masses. However, for temperatures even below the
critical temperature, we overshoot the results from all other lattice
actions.

\subsection{Susceptibilities}
As mentioned above, lattice QCD calculations are largely restricted to
zero chemical potential because of the sign problem. However,
thermodynamic quantities at non-vanishing potentials may be obtained
by Taylor expansion methods as described in Refs.~\cite{Allton:2002zi,
  Ejiri2004, Cheng2009, Schmidt2009}.\par
The pressure at a specific point in the phase diagram $p(T,\mu_B)$
with preferably small chemical potentials is calculated by expanding
the pressure at $p(T,\mu_B = 0)$ around $\mu_B/T$
\begin{equation}
  \label{eq:suscep_quarknumber}
  \frac{p(T,\mu_B)}{T^4} = \sum\limits_{n=0}^{\infty} c_n(T) \left(
    \frac{\mu_B}{T} \right)^n
\end{equation}
with the Taylor coefficients
\begin{equation}
  \label{eq:suscep_cn}
  c_n(T) = \left. \frac{1}{n!} \frac{\partial^n \left( p(T,\mu_B) /
      T^4\right)} {\partial \left( \mu_B / T\right)^n }\right|_{\mu_B = 0}.
\end{equation}
\begin{figure}[t!b]
  \centering 
  \includegraphics[width=1.\columnwidth]{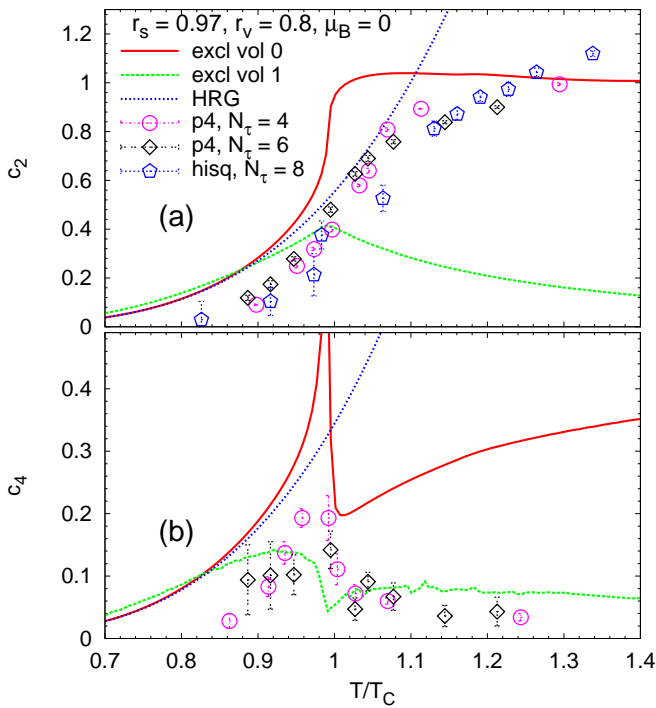}
  \caption{(Color online) Second order (a) and fourth order (b) quark
    number susceptibilities as a function of $T/T_c$ at $\mu_B =
    0$. Results for the interacting hadronic gas ($r_v = 0.8$) with
    (green dashed line) and without (red solid line) excluded volume
    corrections and for the non-interacting ideal hadron resonance gas
    (blue dotted line) are shown compared to lattice
    data~\cite{Cheng2009,Huovinen2010}.}
  \label{fig:sec_four_order_suscep}
\end{figure} 
The coefficients $c_n(T)$ are proportional to the quark number
susceptibilities reflecting quark number
fluctuations~\cite{Karsch2011}.\par
Similarly, by expanding the pressure with respect to the strange
chemical potential $\mu_s$, the first order strange quark
susceptibility $\chi_s(T)$ can be found to be
\begin{equation}
  \label{eq:suscep_strange}
  \chi_s(T) = T^2 \left. \frac{\partial^2 (p(T,\mu_s))}{\partial
      \mu_s^2} \right|_{\mu_s = 0}.
\end{equation}
In contrast to lattice QCD, with our model we are able to calculate
the pressure at any point in the phase diagram and can therefore
numerically calculate the corresponding susceptibilities.\par
\begin{figure}[t!b]
  \centering 
  \includegraphics[width=1.\columnwidth]{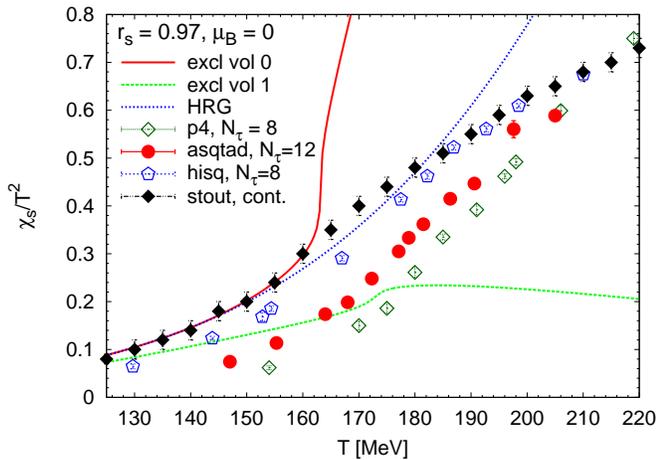}
  \caption{(Color online) Strange quark number susceptibility divided
    by $T^2$ at $\mu_B = 0$ as a function of $T$ for the three
    different cases as described in
    Fig.~\ref{fig:sec_four_order_suscep} together with lattice
    data~\cite{Bazavov2010e, Bazavov2010, Borsanyi2010,
      Aoki2009,Endrodi:2011gv}. The vector coupling strength is set to
    $r_v = 0$.}
  \label{fig:strange_suscep}
\end{figure}

In Fig.~\ref{fig:sec_four_order_suscep} we show the second order $c_2$
(a) and fourth order $c_4$ (b) quark number coefficients in comparison
to lattice QCD data~\cite{Cheng2009, Bazavov2009a, Cheng2010,
  Ejiri2004} at $\mu_B = 0$. Both, the model with and without the
excluded volume effects show a maximum of $c_2$ at the critical
temperature and a smooth decrease for higher temperatures. The fourth
order coefficient $c_4$ exhibits a narrow peak at $T_c$. As expected,
the ideal non-interacting HRG susceptibilities do not show any peak
due to the absence of a phase transition. The susceptibilities of the
ideal HRG come closest to the lattice data with the
p4-action. However, a comparison to the not yet available
susceptibilities from the continuum extrapolated stout action would be
very interesting since the lattice results with the stout action gives
the best agreement for the thermodynamic quantities and the strange
quark susceptibility (Fig.~\ref{fig:strange_suscep}).\par
In our study, we found that the strength of the repulsive vector
coupling has a major influence on the extent of the fluctuations at
the phase transition, even though the derivative is performed at
$\mu_B = 0$. This effect was already pointed out in
Refs.~\cite{Kunihiro1991, Steinheimer2011}. A lower vector repulsion
leads to significantly higher fluctuations of conserved charges at
$T_c$. In our study we set the vector coupling to $r_v = 0.8$ in order
to match our results with lattice data up to $T \approx T_c$. This
vector coupling strength stands in contradiction to the one we found
earlier to be suitable to reproduce a critical end point in the region
suggested by lattice QCD results ($r_v \le 0.2$, cf.\
Fig.~\ref{fig:phase_diagram}).\par
\begin{figure}[t!]
  \centering 
  \includegraphics[width=1.\columnwidth]{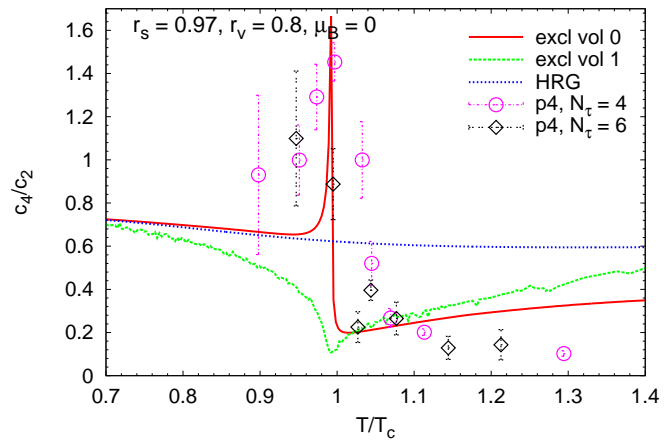}
  \caption{(Color online) Ratio of the fourth order to the second
    order quark number susceptibility as a function of $T/T_c$ at
    $\mu_B = 0$ for the three different cases as described in
    Fig.~\ref{fig:sec_four_order_suscep} together with lattice
    data~~\cite{Cheng2009, Bazavov2009a, Cheng2010, Ejiri2004}.}
  \label{fig:suscep_ratio}
\end{figure}
In Fig.~\ref{fig:suscep_ratio} we show the ratio of the fourth order
to the second order baryon number susceptibility. For the regular
model without finite size corrections the ratio is in line with the
completely flat curve of the HRG up to the critical temperature. The
curve exhibits a discontinuity at $T_c$ with a sudden decrease
above. For higher temperatures the curve is mostly constant around
$c_4/c_2 \approx 0.3$. Due to the effective suppression of degrees of
freedom at the phase transition, the results with excluded volume
effects have a significantly different shape with a much smoother
curve showing only a minimum peak at $T_c$.\par
The strange quark number susceptibility $\chi_s$ divided by $T^2$ is
shown in Fig.~\ref{fig:strange_suscep}. Again, the three cases of our
model yield significantly different results. The curve for the
non-interacting ideal HRG rises again monotonically as expected.  The
regular model with all interactions switched on shows a massive rise
at $T_c$. The results from our interacting model is once again in good
agreement with results from lattice QCD with the continuum
extrapolated stout action up to temperatures of $T_c$. In the case
with the excluded volume effects, in particular heavy strange
particles are suppressed and thus the curve is much lower in this case
showing a maximum at $T_c$ and a slow decrease for higher
temperatures.\par
In a next step we study the susceptibility coefficients at non-zero
baryochemical potentials. Therefore, we compare the coefficients in
three different regions of the phase diagram, namely in the crossover
region, at the critical end point, and at the first order phase
transition. For this purpose, in a first attempt we set the vector
coupling to $r_v = 0$, because this leads to a clearly observable and
well defined phase transition line and a critical end point at
$\mu_B^{\rm crit} \approx 220$~MeV; later on we will also calculate
susceptibilities with more realistic and stronger vector
couplings. Note, that for higher values of $r_v$ this subdivision of
the phase diagram into three clearly distinguishable regions no longer
applies. For a vector coupling strength of $r_v > 0.6$, our model
exhibits only a broad crossover transition in the whole $T$--$\mu_B$
plane (cf.\ Fig.~\ref{fig:phase_diagram}).\par
With our model we then calculate the second and fourth order
susceptibilities along straight lines perpendicular to the phase
boundary going through the points on the phase transition line
$\mu_B^{(1)} = 29.3$~MeV, $T^{(1)} = 161.7$~MeV for the crossover
region, $\mu_B^{(2)} = 216.0$~MeV, $T^{(2)} = 150$~MeV at the critical
end point, and $\mu_B^{(3)} = 489.0$~MeV, $T^{(3)} = 105$~MeV for the
first order phase transition region.\par
\begin{figure}[tb]
  \centering 
  \includegraphics[width=1.\columnwidth]{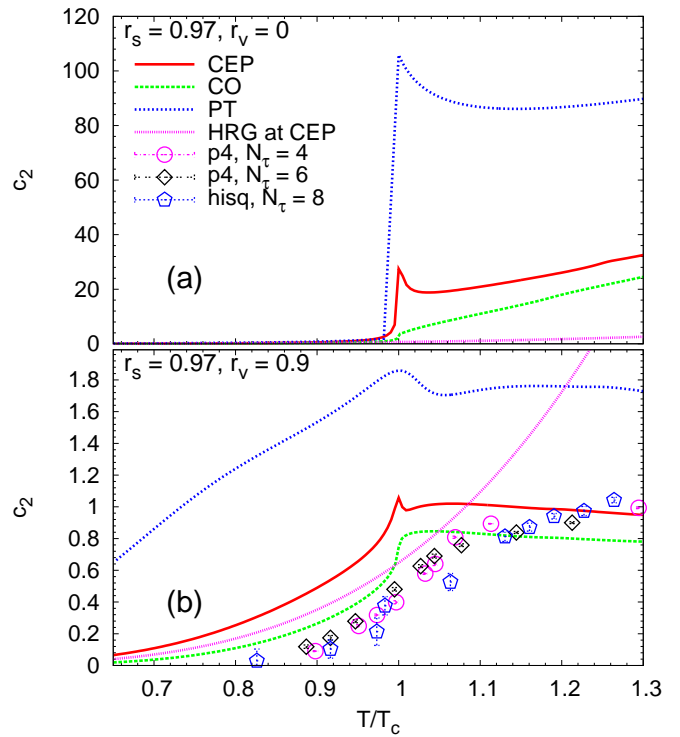}
  \caption{(Color online) Second order quark number susceptibilities
    $c_2$ for vector couplings $r_v = 0$ (a) and $r_v = 0.8$ (b). The
    quantities are calculated going perpendicularly through the phase
    transition line at three different points with $\mu_B \ne 0$. For
    $r_v = 0$ the blue line depicts the crossing of the phase boundary
    in the first order phase transition regime, the green line in the
    crossover regime, and the red line shows $c_2$ going directly
    through the critical end point (see text for more
    information). Shown is also the reference of the ideal hadron
    resonance gas at the critical end point and lattice data at $\mu_B
    = 0$~\cite{Cheng2009, Bazavov2009a, Cheng2010, Ejiri2004} .}
  \label{fig:sec_suscep_pt}
\end{figure}
In Fig.~\ref{fig:sec_suscep_pt} we show the second order coefficients
for the three different regimes for $r_v = 0$ in panel (a) and for
$r_v = 0.9$ in panel (b), together with lattice data~\cite{Cheng2009,
  Bazavov2009a, Cheng2010, Ejiri2004} calculated at vanishing
baryochemical potential $\mu_B = 0$.  Note, that choosing a vanishing
vector coupling of the baryon resonances leads to unreasonably high
susceptibilities at $T_c$ and must be regarded as a limiting test
case. The green dashed curve shows the susceptibilities in the
crossover region at low $\mu_B$.  Due to the smooth transition of the
quantities, the susceptibility only shows a small maximum (rise) at
$T_c$ for $r_v = 0.9$ ($r_v = 0$). The susceptibilities at the
discontinuous first order phase transition (blue dashed line) show the
highest maximum for both values of $r_v$, as expected. At the critical
end point (red solid line), the maximum of $c_2$ is located in between
the results from the crossover and the first order phase transition
region. As shown previously, the susceptibilities for the
non-interacting ideal HRG (pink dashed line) show a monotonous rising
behaviour not being affected by a varied vector coupling strength.\par
\begin{figure}[tb]
  \centering 
  \includegraphics[width=1.\columnwidth]{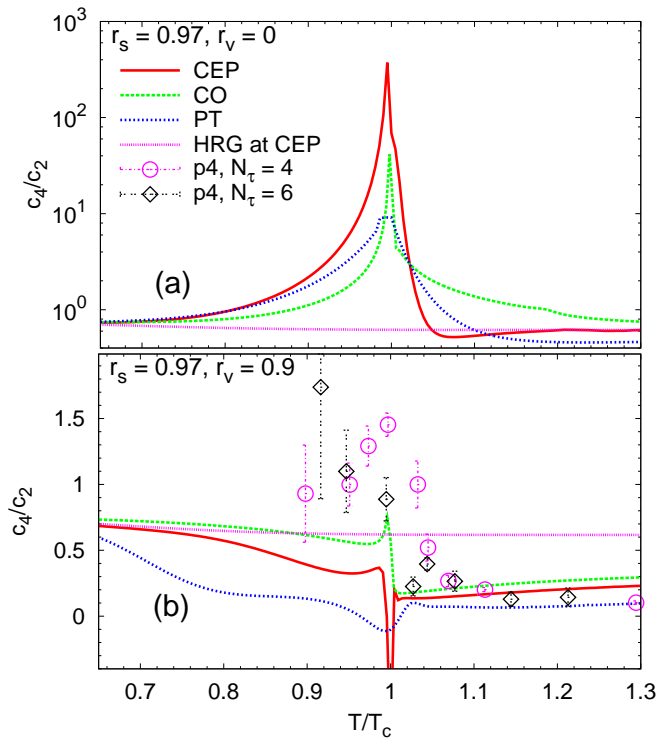}
  \caption{(Color online) Ratios of the fourth order to the second
    order quark number susceptibilities for vector couplings $r_v = 0$
    (a) and $r_v = 0.8$ (b) at different regimes of the phase
    diagram. The line styles are as described in the previous
    figure. Please note the logarithmic scale in the upper panel
    (a).}
  \label{fig:suscep_ratios_at_pt}
\end{figure} 
The ratio of the fourth order to the second order coefficients in the
three different regions of the phase diagram is shown in
Fig.~\ref{fig:suscep_ratios_at_pt} for vector couplings $r_v = 0$ in
panel (a) and $r_v = 0.9$ in panel (b). Once again, the height of the
peak at the critical temperature shows a major dependence on the
strength of the vector coupling. Comparing the curves of the three
different regimes of the phase diagram, the susceptibilities show a
significantly different slope around $T_c$. At non-vanishing
baryochemical potentials the ratios deviate significantly from the HRG
results what is similar to the findings of suppressed fluctuations in
higher order cumulants in the vicinity of the transition region at
zero baryochemical potential~\cite{Friman2011}. In our case the
fluctuations are suppressed by the repulsive vector interactions. This
fact could help to probe the location of the phase transition and a
possible critical end point experimentally by extraction of the
susceptibilities from data.\par
However, the good agreement of our results for the susceptibilities in
the crossover region given a vector coupling strength of $r_v = 0.9$
with the lattice data for vanishing baryochemical potential (green
dashed line in Fig.~\ref{fig:suscep_ratios_at_pt} (b)) together with
the results for the chiral order parameter $\sigma$ in the whole
$T$--$\mu$ plane (Fig.~\ref{fig:phase_diagram}), leads us to the
conclusion, that the existence of a critical end point according to
our model is very questionable. The results from our model suggest
that for reasonable values of the vector coupling, throughout the
entire phase diagram the phase transition is a smooth crossover,
ruling out the existence of a critical end point. This finding
corresponds well with the results from most recent lattice
data~\cite{Endrodi:2011gv}.

\section{Conclusions}
\label{sec:conclusions}

We presented an effective chiral $SU(3)$ model for the QCD equations
of state. In this sigma-omega model we included all known hadrons up
to resonances with masses of $2.6$~GeV together with a parameter $r_v$
that controls the coupling strength of the baryons to the repulsive
vector meson field. Furthermore, we include a finite size effect that
effectively suppresses heavy particles at higher densities.\par
Using this model, at zero baryochemical potential $\mu_B$ we found a
smooth crossover phase transition for the order parameter of the
chiral condensate $\sigma$ with a critical temperature in the range
from $T_c = 164$~MeV to $T_c = 174$~MeV depending on the excluded
volume effects being taken into account. These results are in good
agreement with various data from lattice QCD. Extending this study to
finite $\mu_B$, we show the strong dependence of the phase transition,
i.e.\ position and order, on the vector coupling strength. For
reasonable values of $r_v$, we find that the phase transition is a
smooth crossover in the whole $T$--$\mu_B$ plane and that there is no
critical end point.\par
We also show the thermodynamic quantities from the model and calculate
the quark number and strange quark number susceptibility coefficients
at different values of $\mu_B$. The susceptibilities show a good
qualitatively agreement with lattice data if a sufficiently strong
vector coupling is chosen. This finding underlines the model
suggestion of the non-existence of the critical end point. If the
susceptibilities are extracted at different positions on the phase
boundary, we show that their significantly different behavior may be
used to distinguish the order of the phase transition at a given
point.

\section{Acknowledgements}
\label{sec:ack}

This work was supported by BMBF, GSI, and by the Hessian LOEWE
initiative through the Helmholtz International Center for FAIR (HIC
for FAIR), and the Helmholtz Graduate School for Hadron and Ion
Research (HGS-HIRe). We are grateful to the Center for the Scientific
Computing (CSC) at Frankfurt University for providing the
computational resources.

\end{document}